# Equivalent Circuit Programming for Power Flow Analysis and Optimization

Marko Jereminov, *Member, IEEE* and Larry Pileggi, *Fellow, IEEE*

*Abstract*—The utility of domain-specific knowledge for modeling, simulation, and optimization has been demonstrated for various research problem domains, including power systems. The concept of Equivalent Circuit Programming was previously developed and facilitated for robust, efficient, and scalable solution of network simulation and optimization problems. This paper extends the theoretical foundation of Equivalent Circuit Programming to enable the fusion of optimization theory and algorithms with the numerical methods that utilize the domain-specific knowledge of power flow models. The generality, scalability, and numerical robustness of the resulting framework are demonstrated on realistic AC power flow (ACPF) models of up to 70k buses with proper enforcement of industry-required operational and security constraints.

## I. Numerical Methods taught by the First Principles

Optimal decision making represents a major component of everyday life and can be found everywhere, especially in emerging technologies increasingly dependent on Machine Learning algorithms and Artificial Intelligence [1]-[2]. The process of efficiently, accurately, and robustly obtaining these optimal decisions, however, can be extremely challenging since the closed form analytical solutions usually do not exist [1]. Therefore, developing numerical methods to accurately determine optimal decisions has become one of the most prominent problems since the early days of research in operations science and mathematical optimizations.

To some extent, most of the developed numerical algorithms [1]-[3] mimic and utilize the knowledge of natural optimization processes that continuously occur in the world around us. The gradient based methods [1] use the idea that the fastest way to go down the hill, i.e., reach an optimal solution, is to take the steepest descent. Furthermore, metaheuristics such as the Simulated Annealing algorithm [4]-[5] mimic the natural processes that are found in the heating and controlled cooling in metallurgy, while the Evolutionary algorithms (EA) [4],[6] use the idea of biological evolution and survival of the fittest as a base for finding the global optimal solution of a mathematical optimization problem.

Interestingly, as it is the case in natural optimization processes where the "solution methodology" depends on a law of physics that governs the process, the fact that there doesn't exist an algorithm that can efficiently work for all of the mathematical optimization problems is demonstrated by so-called "No Free Lunch" (NFL) theorems in [7]. By examining the connection between the effective algorithms and problems they are solving, the authors from [7] proved that if no domain-specific knowledge of a problem is considered within the algorithm, all the algorithms should perform the same on average once applied to the complete spectrum of problems. Therefore, it becomes apparent that to obtain a more efficient and numerically robust methodology for solving a particular class of optimization problems, all the known domain specific information must be taken into account, particularly when dealing with the optimization of physical systems.

Long before the formal proof and introduction of NFL theorems, utilizing the domain specific knowledge was already shown to facilitate powerful theorem proofs and solution methodologies. Namely, one of the main theorems that defines the conservation of energy within a network, Tellegen's Theorem [8], was proven by the Kirchhoff Current and Voltage Laws (KCL and KVL). Moreover, the electronic circuit simulator SPICE [9] and its many derivatives [10]-[12] have utilized domain-specific knowledge of transistors and other circuit elements to enable the simulation of the highly nonlinear circuit problems with billions of variables [13].

Inspired by the circuit theoretic approach to solving simulation problems and backed by NFL theorems, the Equivalent Circuit Programming (ECP) [14] framework was introduced and shown to facilitate robust, scalable, and efficient numerical solution processes of network analyses. It was shown that the optimality conditions of a network optimization problem can be fully and accurately represented in terms of equivalent circuit models and respective variables [11],[15]. This perspective on the problem optimality conditions facilitated embedding of domain-specific knowledge directly within the optimization theory and corresponding state-of-the-art numerical methods [1],[3],[16]-[18] to develop a set of novel algorithms that are infused by network physics. Importantly, the complete consideration of the domain-specific knowledge enabled the merging of decades of advanced research in the circuit simulation community together with the advances in state-of-the-art optimization techniques and algorithms. This,

This work was supported in part by the Defense Advanced Research Projects Agency for the RADICS Program under Award FA8750-17-1-0059, and in part by the National Science Foundation under Contract ECCS-1800812.

Marko Jereminov is with Pearl Street Technologies, Pittsburgh, PA, 15206 USA (e-mail: mjereminov@pearlstreettechnologies.com).

Larry Pileggi is with the Dept. of Elect. and Comp. Eng., Carnegie Mellon University, Pittsburgh, PA 15213 USA (e-mail: pileggi@andrew.cmu.edu).

as NFL theorems proved, demonstrated that the *network physics does not necessarily make the simulation and optimization problems harder, but on the contrary, can help in achieving more efficient, and scalable framework, if considered and utilized within the state-of-the-art optimization theory and algorithms*.

As a cornerstone of economic health and an essential service in modern societies [19], an electrical power grid is governed by physical conservation laws (KCL and KVL) that define the relationship between the currents and voltages within the system. Recently, power grid steady-state analyses [20]-[23] have included an equivalent circuit formulation for representing the power flow problem in terms of current, voltage and admittance state variables [24]. It was shown that this approach enables more stable and robust numerical convergence properties [23]-[26] once compared to traditionally used formulations [27]-[29]. The equivalent circuit formalism set a foundation for incorporation of power flow models within the novel ECP framework presented in this paper.

To motivate the extension and application of Equivalent Circuit Programming theory and algorithms, this paper builds on the concept of physics-inspired numerical methods to address impediments and combine the advantages of traditional numerical algorithms, and the recently introduced circuit simulation approach for solving a wide scope of AC power flow problems. To that end, the major contributions of this paper are:

i.  Application of ECP theory to include an AC power flow model and introduce a novel perspective on representing and analyzing its set of optimality conditions and duality theory in terms of conservation of energy within the system.

ii. Development of a set of ECP algorithms that builds on the ECP theory to combine the advantages of both optimization and physics-based circuit simulation approaches by embedding the power system domain-specific knowledge within the generic optimization techniques and algorithms.

The efficacy, robustness, and scalability that can be achieved within the introduced framework is demonstrated by analyzing a set of realistic power flow models of up to 70k buses with industry required operational constraints enforced. The considered analyses include, but are not limited to, realistic AC power flow (ACPF) simulations, as well as AC optimal power flow (OPF) and security constrained AC-OPF analyses.

## II. FORMULATING AND SOLVING AN AC POWER FLOW MODEL

Characterizing and analyzing a power grid steady-state response in terms of power-mismatch formulation and phasor voltage state variables [28] has been widely accepted as a standard, particularly for transmission level grid analyses. However, as any other electrical circuit, an electrical power system is governed by physical conservation laws that are naturally defined in terms of current and voltage state variables. Notably, as found in the publications dating back to 1940s and 1950s [30]-[32], the original formulations for analyzing the grid response utilized these classical state variables, and was the first formulation implemented on a digital computer by Ward and Hale in 1956 [31]. These initial current/voltage-based formulations, however, suffered from serious drawbacks in efficiency, accuracy, and scalability [33]-[35], hence leading power system analyses to the power mismatch formulation for characterizing the network steady-state response.

Recent advances in power flow modeling and analysis [22], however, have demonstrated that the ACPF problem can be modeled and solved as a traditional circuit simulation problem represented in terms of KCL and KVL equations [20]. The formalism introduced by the equivalent circuit representation of an ACPF problem enabled the application of circuit simulation techniques and homotopy methods to improve the simulation robustness and scalability [23]-[25].

### A. AC Power Flow equivalent circuit model

Consider a power system whose steady-state response, and hence its power flow solution, is characterized in terms of fundamental frequency phasor voltages and currents ($\bar{V}_i = V_{R,i} + jV_{I,i}$ and $\bar{I}_i = I_{R,i} + jI_{I,i}$), with the relationships between a set of generators $\mathcal{G}_B$ and load demands $\mathcal{D}_B$, interconnected by a set of transmission network elements, $\mathcal{T}_X$. Next we outline how the equivalent circuit modeling of the power flow problem can overcome the challenges of current-voltage based formulations [30]-[31],[35].

#### 1) Linear circuit elements: Bus admittance matrix

For the "Power-Mismatch" (P/Q) formulations [28], the network elements, such as the ones presented in Figure 1, correspond to the main source of inherent nonlinearities in the power flow *bus admittance matrix* ($Y_{BUS}$). In contrast, for the "Current-Mismatch" (I/V) formulations, the admittance matrix corresponds to linear constraints relating phasor currents and voltages, as governed by Ohm's Law. The main source of nonlinearities now shifts towards enforcing the generator/load power-based operational constraints.

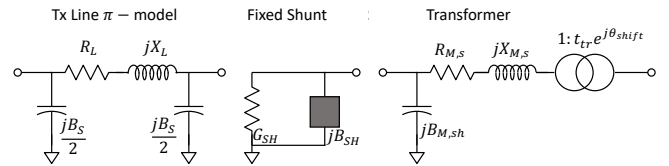

Figure 1. Example of naturally linear power flow impedance models.

#### 2) Modeling constant power elements: GB bus model

Characterizing the two most common transmission system components, PV and PQ power flow models (Figure 2), in terms of phasor current and voltage state variables represents the main impediment [32]-[33] for the original "Current-Voltage" formulations in terms of accuracy [31]-[32] and robustness [33].

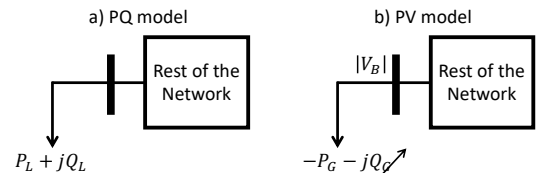

Figure 2. Power models in an ACPF problem. Constant PQ model (a) absorbs the real and reactive powers ($P_L$ and $Q_L$) from the network. On the other side, in addition to constant real power ($P_G$) supplied to the network, the PV model (b), adjusts the reactive power ($Q_G$) to maintain the voltage magnitude $|V_B|$ of a system bus.

To circumvent the challenge of relating the phasor current/voltage states with the power-based operational constraints, the equivalent circuit modeling approach introduces an additional set of admittance state variables [14], [24]. Representing PQ and PV buses in terms of generalized GB bus model (Figure 3), *power-mismatch constraints are enforced by solving for the values of admittance variables that ensure that the power injections to the network, modeled by the bus admittance matrix, match the pre-specified power constraints,* as given for $i^{th}$ bus in (3)-(4). Most importantly, the PQ-based ACPF models are now included within a set of current mismatch equations (1)-(2) in terms of *unknown* admittances that constrain the relationship between phasor currents and voltages, while corresponding to nonlinearities within the formulation:

$$V_{R,i}: \ G_i V_{R,i} - B_i V_{I,i} + \text{Re}\{Y_{BUS}\widetilde{\boldsymbol{V}}\}_i = 0 \tag{1}$$

$$V_{I,i}: \ G_i V_{I,i} + B_i V_{R,i} + \text{Im}\{Y_{BUS}\widetilde{\boldsymbol{V}}\}_i = 0 \tag{2}$$

$$G_i: \ G_i\left(V_{R,i}^2 + V_{I,i}^2\right) = \sum_{k=1}^{|\mathcal{D}_B|} P_{L,k} - \sum_{m=1}^{|\mathcal{G}_B|} P_{G,m} \tag{3}$$

$$B_i: \ B_i\left(V_{R,i}^2 + V_{I,i}^2\right) - Q_{G,i} = -\sum_{k=1}^{|\mathcal{D}_B|} Q_{L,k} \tag{4}$$

In addition to ensuring the hold of power mismatch conditions on a set of KCL equations (1)-(2), the introduction of admittance state variables and corresponding power-mismatch equations (3)-(4) provides an additional degree of freedom for enforcing other operational constraints. Namely, constraints given in terms of system currents and voltages (thermal current limits, switch shunt control, etc.) more naturally fit, and hence can be more readily enforced, within the set of KCL equations (1)-(2). On the other side, operational constraints and control defined in terms of real and reactive powers can be included within a set of newly added GB equations (3)-(4). For instance, the voltage regulation (VR) characteristics of a PV model, given in Figure 4, is enabled by introducing an additional reactive power variable, $Q_{G,i}$ to the respective equation as shown in (4), for which the additional bus voltage control constraints are enforced as:

$$Q_{G,i}: \ V_{R,i}^2 + V_{I,i}^2 - \Delta v^+ + \Delta v^- = |V_s|^2 \tag{5}$$

$$\Delta v^+: \ \Delta v^+(Q_{MAX} - Q_G) \to 0^+ \tag{6}$$

$$\Delta v^-: \ \Delta v^-(Q_G - Q_{MIN}) \to 0^+$$

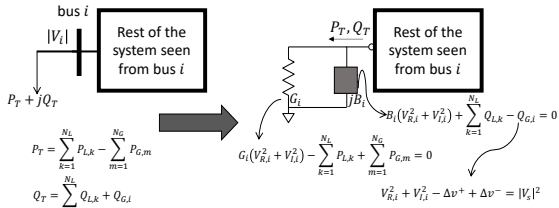

Figure 3. GB bus representation of a constant power model.

Finally, the complete set of circuit equations $F_{ckt}$ and state variables $\boldsymbol{x}$ represented per bus as in (1)-(6) are defined as in (7), and numerically solved to obtain an ACPF solution.

$$F_{ckt}(\boldsymbol{x}) = \boldsymbol{0} \tag{7}$$

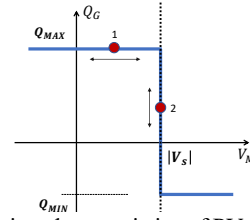

Figure 4. Disjunctive characteristics of PV voltage regulation. If reactive power $Q_{G,i}$ doesn't approach operational limits (point 2) the voltage is maintained at a set point ($|V_s^k|$), while whenever one of the limits is reached (point 1), the regulation is not maintained, and the controlled setpoint is relaxed in directions given by $\Delta v^+$ and $\Delta v^-$.

## 3. Numerical robustness: A circuit simulation problem

Robustness and efficacy of numerical methods employed for solving a system of nonlinear power flow equations are challenges for both traditional "Power-Mismatch" [27] and "Current-Mismatch" [31],[35] formulations. Consider a Newton Raphson (NR) method that utilizes the first order Taylor expansion for iterative improvements, $\Delta \boldsymbol{x} \in \mathbb{R}^{|F_{ckt}(x)|}$, from the initial starting point toward a solution of the nonlinear system (7), as shown in Figure 5. In relating a particular ACPF problem reformulation with the convergence properties of NR method, the lessons learned from deeply developed circuit simulation methods [9]-[12] indicate that the control of the NR-step size is the difference between convergence and divergence of the iterative process.

Figure 5. NR process of solving a circuit simulation problem.

The main difference between a generalized NR method and a circuit simulation-based formulation is the utilization of domain-specific knowledge to control the iterative process. Contrary to the no-step ($\boldsymbol{T} = 1$) control, or a constant NR-step damping ($\boldsymbol{T} \in \mathbb{R}^1$) employed in generic numerical toolboxes [1],[3], circuit simulation techniques limit each term of the NR-step ($\boldsymbol{T} \in \mathbb{R}^{|\Delta x|}$) based on the understanding of equivalent circuit physical characteristics. Most importantly, even though this kind of vectorized NR step damping *doesn't* ensure that the residual of the problem is decreased at every iteration, the approximated "trust region" techniques based on the problem "physics" have been shown to work quite effectively, particularly once combined with homotopy methods [25]-[26].

### B. Solving an ACPF constrained optimization problem

With the introduced equivalent circuit representation of an ACPF model, we next focus on the framework for optimal building (design) of its set-points and control parameters. Therefore, consider a generic representation of an ACPF constrained optimization problem with a continuously differentiable objective function $\mathcal{F}(\boldsymbol{x})$ and a set $C_E(\boldsymbol{x})$ of power flow operational $F_{ckt}(\boldsymbol{x})$ constraints and limits $h(\boldsymbol{x})$:

$$\min_{\boldsymbol{x} \in C_E} \mathcal{F}(\boldsymbol{x})$$
$$C_E = \{\boldsymbol{x} \mid F_{ckt}(\boldsymbol{x}) = 0, h(\boldsymbol{x}) < 0\} \quad (8)$$

To enforce the domain of a given set $C_E$ onto the minimization objective from (8), the objective function is augmented with a weighted sum of constraints within a Lagrangian function:

$$\mathcal{L}(\boldsymbol{x}, \boldsymbol{\lambda}, \boldsymbol{\mu}) = \mathcal{F}(\boldsymbol{x}) + \boldsymbol{\lambda}^T F_{ckt}(\boldsymbol{x}) + \boldsymbol{\mu}^T h(\boldsymbol{x}) \quad (9)$$

where the vectors $\boldsymbol{\lambda} \in \mathbb{R}^{|F_{ckt}(\boldsymbol{x})|}$ and $\boldsymbol{\mu} \in \mathbb{R}^{|h(\boldsymbol{x})|}$ represent the "optimal weights" or dual variables related to power flow operational constraints and limits respectively.

If it exists, an operationally constrained power flow solution that minimizes the Lagrangian function in (9), and is given by a stationary point $\boldsymbol{z} = [\boldsymbol{x}, \boldsymbol{\lambda}, \boldsymbol{\mu}]^T$, can be obtained from a set of dual, primal and complementarity conditions derived from the first order differentiation of the Lagrangian function as:

$$\theta(\boldsymbol{z}) = \begin{cases} \nabla_x \mathcal{F}(\boldsymbol{x}) + \nabla_x F_{ckt}(\boldsymbol{x})^T \boldsymbol{\lambda} + \nabla_x h(\boldsymbol{x})^T \boldsymbol{\mu} \\ F_{ckt}(\boldsymbol{x}) \\ \boldsymbol{\mu} \circ h(\boldsymbol{x}) + \boldsymbol{\varepsilon} \end{cases} = \boldsymbol{0} \quad (10)$$

with $\varepsilon \to 0^+$ to ensure $\theta(\boldsymbol{z})$ differentiability, and under the consideration of primal (11) and dual (12) feasibility:

$$\boldsymbol{\mu} \geqslant \boldsymbol{0} \quad (11)$$
$$h(\boldsymbol{x}) \leqslant \boldsymbol{0} \quad (12)$$

Like in the NR method defined in Figure 5, an operationally constrained power flow solution can be obtained by linearizing and iteratively solving the set of equations from (10), while controlling a NR-step $\Delta \boldsymbol{z}$ and complementarity violation coefficient $\varepsilon$ as presented in Algorithm 1. Importantly, interior point methods (IPMs) are the most popular numerical methods for solving the aforementioned optimization. However, in contrast to solving a circuit simulation problem, IPMs utilize a constant NR-step damping ($T = t_{pd}$) that ensures residual decrement (if feasible) at every iteration. Moreover, other versions of IPM mostly differ in the way of handling $\varepsilon$ control [1]-[3],[16]-[18], while utilizing constant NR-step damping.

---

**Algorithm 1.** *Short-step (Primal-Dual) Interior Point Method (IPM)*.
**Given**: feasible $\boldsymbol{z^0}$, and tolerance $\epsilon > 0$, $\alpha \in (0,0.5)$ and $\beta, \sigma \in (0,1)$
**Repeat**:
1. Set $\varepsilon = -\sigma h(\boldsymbol{x^k})^T \boldsymbol{\mu^k}$ and Compute NR step $\Delta \boldsymbol{z}$
2. **Apply a form of line search on** $\|\theta(\boldsymbol{z^k} + t_{pd}\Delta \boldsymbol{z})\|_2$
   a. Compute a constant, $t_{pd}^{max}$ that ensures dual feasibility:
   $$t_{pd}^{max} = 0.99 \min\left\{1, \min\left\{-\frac{\mu_i^k}{\Delta \mu_i} \mid \Delta \mu_i < 0, \forall i \in [1, |\mu|]\right\}\right\}$$
   b. Starting from $t_{pd} \to t_{pd}^{max}$ continue damping to ensure:
   $$h(\boldsymbol{x^k} + t_{pd}\Delta \boldsymbol{x}) < 0$$
   c. Solve line search [1] to satisfy a form of residual condition
3. Update NR step: $\boldsymbol{z^{k+1}} = \boldsymbol{z^k} + t_{pd}\Delta \boldsymbol{z}$, and increase: $k \to k+1$
**Until**: $\|\theta(\boldsymbol{z^{k+1}})\|_2 \leq \epsilon$ and $\varepsilon \leq \epsilon$

---

It must be noted that the operationally constrained power flow solution can be called *optimal* only if the Second-Order necessary condition from (13) holds. That is, the Hessian of the Lagrangian function, $\nabla_{xx}^2 \mathcal{L}(*)$ evaluated at that point must be positive-definite on a feasible step-size perturbation $\boldsymbol{\tau}$, with $T_{X^*}$ representing the tangent linear sub-space at $\boldsymbol{x^*}$ [1]-[3].

$$\boldsymbol{\tau}^T \nabla_{xx}^2 \mathcal{L}(\boldsymbol{x^*}, \boldsymbol{\lambda^*}, \boldsymbol{\mu^*})\boldsymbol{\tau} > 0, \forall (\boldsymbol{\tau} \neq \boldsymbol{0}) \in T_{X^*} \quad (13)$$

### C. Optimization vs circuit simulation approach

Solving a nonlinear constrained optimization problem, such as optimizing an AC power flow model with consideration of the industry required operational constraints and control limits (8), can be a very challenging task that is prone to divergence, very slow convergence, or convergence to nonoptimal saddle points [1]. Moreover, when solved using commercial optimization toolboxes, the optimization problem relies on careful tuning of the solver parameters [1].

From the perspective of NR step control, these challenges can be traced to a constant NR step damping parameter obtained as a solution to some form of a line search problem. Namely:

1. The line search may not have a feasible solution due to the problem nonlinearities or bad initial starting point. Hence, only the increase of residual can allow for the future convergence of the iterative process.
2. The introduction of complementarity constraints requires the additional step-size damping to ensure the primal and dual feasibility. This can cause scalability issues, since one step size can possibly saturate the whole solution vector.
3. The unnecessary damping of certain variables can force the iteration process to remain stuck in a local area, thereby increasing the chances of converging to a local saddle point, as depicted in Figure 6.

Conversely, circuit simulation algorithms employ the knowledge of the problem's physical characteristics to limit NR step size, which means that each variable is treated and limited separately. While *this may not necessarily decrease the residual at every iteration*, it can be beneficial or problematic. If not limited properly, the convergence process can take a step from which it cannot recover, causing the future divergence. To this end, *this paper focuses on merging the best from the two approaches* and developing a set of techniques that can potentially improve scalability and efficiency by ensuring that the vectorized damping of a NR step also results in the optimal residual decrement at each iteration.

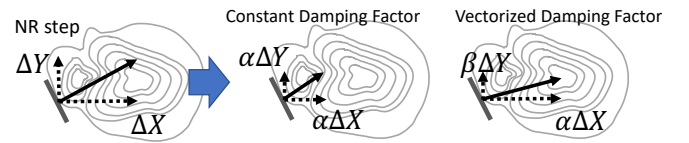

Figure 6. Drawbacks of a single step limiting factor $\alpha$.

### III. EQUIVALENT CIRCUIT PROGRAMMING THEORY

Equivalent Circuit Programming (ECP) was introduced [14] as an approach that is inspired by circuit simulation concepts and NFL (no free lunch) theorems. ECP transforms the "art of tuning" [1] nonlinear algorithms into a technology that utilizes the domain specific knowledge of network optimization problems to ensure the numerical efficacy and scalability. To specifically develop a set of physics-inspired numerical algorithms for ACPF analyses, this section provides a perspective on interpreting the duality theory and optimality conditions of an ACPF constrained problem *in terms of conservation of energy within a system*.

### A. Necessary optimality conditions as a circuit model

#### 1. Primal problem ($F_{ckt}(x) = 0$)

As the first component of the Necessary conditions from (10), the primal problem itself corresponds to a set of equations governing an equivalent circuit model (7).

#### 2. Dual problem ($\nabla_x \mathcal{F}(x) + \nabla_x F_{ckt}(x)^T \lambda + \nabla_x h(x)^T \mu = 0$)

Different reformulations of a problem correspond to different duals [1]. Therefore, it is expected that following first principles to a physical system as an equivalent circuit, also maps to a network model in dual space. To that end, the ECP theory [14] generalizes the adjoint network concept previously used for noise analysis in Radiofrequency (RF) circuits [37]. Accordingly, the dual space mapping of elements defining ACPF models, presented in Table 1, is obtained by analyzing [14] the effect of admittance perturbations on the conservation of power within the network.

Table 1: Mapping from an original to the adjoint network element.

| Original Network | | Adjoint Network |
|---|---|---|
| Independent current source | → | Open circuit |
| Independent voltage source | → | Short circuit |
| Capacitive/inductive impedance | → | Inductive/Capacitive impedance |
| Voltage step-up/down transformer | → | Current step-up/down transformer |
| Current step-up/down transformer | → | Voltage step-down/up transformer |

For instance, if an optimization constraint is given in terms of Ohm's Law, which relates a current and voltage of an inductive impedance, its dual space equivalent is represented by the Ohm's Law governing the current and voltage of a capacitive impedance. The adjoint current and voltage then uniquely correspond to the Lagrange multipliers of the respective optimization constraint. With each of the equivalent circuit elements mapped to its respective adjoint (dual representation), the resulting set of adjoint circuit equations then corresponds to the dual problem contributions from a set of network constraints, $\nabla_x F_{ckt}(x)^T \lambda$. When combined with the contributions from the optimization objective $\nabla_x \mathcal{F}(x)$, and a set of enforced operational limits, $\nabla_x h(x)^T \mu$, it results in a dual problem as given in (10). Our network representation of the ACPF dual problem will consider the adjoint network model for the remaining discussions and contributions.

##### a. Adjoint network model for representing the First order ACPF sensitivity, $\nabla_x F_{ckt}(x)^T \lambda$.

To obtain the physical system representation of a dual problem from (10), consider a set of equivalent circuit models representing an ACPF problem as introduced in Section II-A. For a given set of models defined by the bus admittance matrix (Figure 1), the respective dual space representations are obtained by following the mappings from Table 1. Namely, any impedance is represented by its complex conjugate, while a voltage transformer maps to a current transformer with the inverse complex turns ratio, as shown in Figure 7. Similarly, the same rule follows for variable admittances introduced to enforce power constraints on the governing circuit equations.

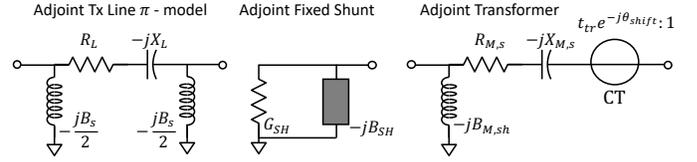

Figure 7. ACPF admittance models from Figure 1 in dual space.

##### b. Active constraints and objective function gradient $\nabla_x \mathcal{F}(x)$, as excitation sources of the adjoint network

The ECP theory in [14] showed that the excitation of the adjoint network corresponds to the first-order sensitivities of the problem objective, $\nabla_x \mathcal{F}(x)$, and a set of active operational limits. Hence, an adjoint network characterizes the first-order sensitivities of original ACPF model and further "connects" them with the sensitivities of problem objective function, and an active set of operational limits. This results in a physical perspective and representation of the dual problem (10) in terms of the laws of conservation of energy within the adjoint (dual) system. It follows that an optimal solution candidate to the problem represents an operating point at which the power supplied by the dual excitation sources (objective sensitivities, active operational limits) matches the power absorbed by the adjoint network (ACPF sensitivities).

#### 3. Complementarity conditions ($\mu \circ h(x) = -\varepsilon$)

A physical intuition behind a set of complementarity conditions that can be utilized within ECP algorithms can be obtained by noting that the complementary "switch-like" characteristics resemble diodes in electronic circuit problems, as shown in Figure 8. Roughly speaking, after a threshold voltage point (diode thermal voltage, $V_{th}$) across a diode is reached, the conducting current ($I_D$) flows with exponential growth as a function of that voltage. Similarly, as the network state approaches its operational limit, a dual variable $\mu$ "activates" and becomes nonzero. Therefore, inspired by the circuit simulation for handling diode nonlinearities, and in contrast to traditionally used IPMs, in ECP we propose to directly fix the value of the non-ideality factor $\varepsilon$ to a preset small value ranging from $10^{-6}$ to $10^{-12}$ [14]. Hence, a set of ECP NR-step limiting techniques is developed for handling such nonlinear characteristics within the NR method.

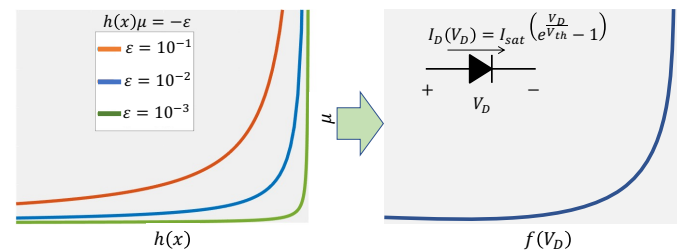

Figure 8. A diode circuit perspective for understanding and future handling the complementarity conditions within ECP algorithms.

### B. Passivity of adjoint (dual) network ensures optimality

From the perspective of conservation of energy within a system, a solution to a set of necessary optimality conditions from (10) represents an operationally constrained ACPF solution for which the power supplied by the dual excitation sources (e.g., objective sensitivities) matches the power absorbed by the adjoint network. The ECP theory proves in [14] that such a point is said to be an *optimal* operating point if the adjoint network (dual problem) remains *passive* [36] around its operating point. Namely, a change in the total power absorbed by the adjoint network due to the small system perturbation is positive, i.e., there is no point at which the adjoint network starts to "generate power." The passivity condition is a well-known concept in circuit theory and is now utilized together with the physical representation of the necessary optimality conditions in building a set of novel ECP algorithms.

## IV. EQUIVALENT CIRCUIT PROGRAMMING ALGORITHMS

Besides the traditional approaches for solving an ACPF optimization problem, optimizing an ACPF model also corresponds to *simulating a network and its additionally excited adjoint model, whose passivity at the operating point further guarantees its optimality.* Building on this new perspective, and the advantages of traditional optimization techniques, this section introduces a set of novel techniques for NR step control. In referring to Figure 5 and Algorithm 1, methods for handling the NR step control are replaced by a set of new techniques presented in Algorithm 2.

---

**Algorithm 2. ECP approach for vectorized NR step limiting**
**Input:** network model of optimality conditions and NR step, $\Delta z$
**APPLY**
    **1. Diode limiting techniques**
      • *Static diode limiting* (physics/optimization inspired)
        ○ $\Delta z \rightarrow \mathbf{T_{SDL}} \circ \Delta z$
      • *Dynamic diode limiting* (physics inspired)
        ○ $\Delta z \rightarrow \mathbf{T_{DDL}} \circ \Delta z$
    **2. Optimal residual limiting** (optimization inspired)
    **IF:** $\|\theta(z^k + \mathbf{T_{ORL}} \circ \Delta z)\|_2^2 \approx \|\theta(z^k)\|_2^2 \nrightarrow 0$
      • *Variable filtering* (physics inspired) **GOTO: Step 2**
    **ELSE:** $\Delta z \rightarrow \mathbf{T_{ORL}} \circ \Delta z$
**Output:** Controlled NR-step $\Delta z$

---

While diode limiting techniques ensure the primal (11) and dual (12) feasibility by handling the steepness of complementarity conditions, variable filtering can be derived via physics-inspired trust regions based on physical representation of (10). In contrast, optimal residual limiting enables the residual decrement at each iteration, thereby overcoming the impediments of traditional circuit simulation approaches. Notably, the introduced set of vectorized damping techniques considers and limits each variable within NR step separately, while at the same time decreases the problem residual. Most importantly, techniques presented in Algorithm 2 are generic and applicable to other ACPF formulations within existing simulation and optimization frameworks.

### A. Diode Limiting Techniques

A set of ACPF operational limits, $h(x)$, and respective complementarity conditions can be further defined in terms of

upper and lower bounds ($X_U$ and $X_L$) on the state variables of equivalent circuit model as:

$$h(x) = \{x \mid X_L - x \leq 0, x - X_U \leq 0\} \tag{14}$$

$$\mu \circ h(x) = \kappa(\mu, x) \equiv \begin{bmatrix} \mu_L \circ (X_L - x) \\ \mu_U \circ (x - X_U) \end{bmatrix} = -\varepsilon \tag{15}$$

where $\mu_L$ and $\mu_U$ represent dual variables related to subsets of lower and upper operational limits respectively from (14).

To ensure that consecutive NR-steps remain feasible and well controlled within the steep nonlinear complementarity curve (e.g. Figure 8), our proposed diode limiting technique combines concepts relating to the primal/dual feasibility step damping (*this time per variable*) from Algorithm 1, and circuit simulation diode damping approaches in [11]. We describe this new approach in the following subsection.

#### 1. Static Diode limiting technique

Consider a state variable, $x_i$ bounded by its upper ($X_{U,i}$) and lower ($X_{L,i}$) operational limits. Furthermore, let $\mu_{B \in \{U, L\}, i}$ correspond to the respective pair of dual variables, lower bounded by a feasibility condition given in terms of $\varepsilon$-factor:

$$\mu_{B,i} > \mu_{min,i} = \frac{\varepsilon}{X_{U,i} - X_{L,i}} \rightarrow 0^+ \tag{16}$$

A pair of NR step damping factors ($\tau_{x,i}$, $\tau_{B \in \{U, L\}, i} \in [0,1]$) that ensures the primal and dual NR step feasibility; i.e. $\Delta \bar{x}_i \rightarrow \tau_{x,i} \Delta x_i$ and $\Delta \bar{\mu}_{B,i} \rightarrow \tau_{B,i} \Delta \mu_{B,i}$, is then obtained as:

$$\tau_{x,i} = \min[1, \alpha_{DFS} \Delta_X] \tag{17}$$

$$\tau_{B,i} = \begin{cases} \min\left[1, \alpha_{DFS}\left(\frac{\mu_{min,i}}{\Delta \mu_{B,i}} - \Delta_\mu\right)\right] & \text{if } \Delta \mu_{B,i} < 0 \\ 1 & \text{else} \end{cases} \tag{18}$$

with $\alpha_{DFS} = 0.99$, introduced to prevent numerical noise issues, while the damping ratios $\Delta_X$ and $\Delta_B$ are given by (19)-(20). Moreover, additional to separately limiting each variable NR step, the purely circuit simulation approach often manually controls the value of $\alpha_{DFS}$ as a tuning parameter to maintain conditioning of the iterative process. Hence, to remove the possible negative effects that parameter tuning may introduce, the second "diode" limiting ensures that the conditioning of iterative process becomes "controlled" by the problem itself.

$$\Delta_X = \begin{cases} \dfrac{X_{U,i} - x_i^k}{\Delta x_i} & \text{if } \Delta x_i > 0 \\ \dfrac{X_{L,i} - x_i^k}{\Delta x_i} & \text{if } \Delta x_i < 0 \end{cases} \tag{19}$$

$$\Delta_B = \frac{\mu_{B,i}^k}{\Delta \mu_{B,i}} \tag{20}$$

#### 2. Dynamic Diode limiting technique

For maintaining the convergence stability, Dynamical Diode limiting further damps ($\delta_{x,i}, \delta_{B,i} \in [0,1]$) the feasible NR-steps to ensure that the complementarity residual converges concurrently with the reminder of the problem. This is achieved by controlling the rate of change of the complementarity function residuals between the consecutive NR iterations as:

$$\frac{\kappa\left(\mu_{B,i}^k + \delta_{B,i}\Delta\bar{\mu}_{B,i}, x_i^k + \delta_{x,i}\Delta\bar{x}_i\right)}{\kappa\left(\mu_{B,i}^k, x_i^k\right)} \geq 1 - \rho \quad (21)$$

with $\rho$ being a function of normalized residual $\chi^k$, given by (22) and Figure 9 that dynamically controls the rate of change.

$$\rho = \rho(\chi^k) = \begin{cases} \dfrac{\log\chi^k}{\sqrt{\chi^k}} & \text{if } \chi^k > e^2 \\ 2e^{-1} & \text{else} \end{cases} \quad (22)$$

$$\chi^k = \max\left[e^{\frac{31}{10}}, \kappa(x^0, \mu_B^0)^{-1}\right]\frac{\|\theta(\mathbf{z}^k)\|_2^2}{\|\theta(\mathbf{z}^0)\|_2^2} \quad (23)$$

In referring to Figure 9, the smaller initial complementarity residual is, the smaller allowable initial decrement $\rho(\chi^0)$, indicating that the complementary variables are in the steep region and need to be damped more. Moreover, as the $\|\theta(\mathbf{z}^k)\|_2^2$ converges, the condition from (21) relaxes, thereby requiring less step damping towards the overall problem convergence.

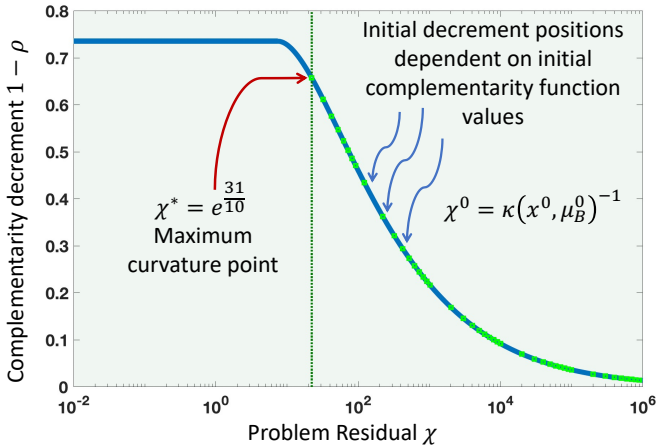

Figure 9. Dynamic damping function representing an amount of complementarity decrement as governed by the problem residuals.

Next, consider the system of two conditions from (21) given for upper and lower ($B \in \{U, L\}$) complementarity constraints. The residual dependent step limiting factors in Algorithm 3 is then computed from the obtained system of inequalities.

---

**Algorithm 3. Dynamic Diode limiting damping factors calculation**

**Input:** $\Delta\bar{x}_i, \Delta\bar{\mu}_{B,i}$ and damping ratios $\Delta_{XB}, \Delta_{\rho B}, \Delta_{\delta B}$ given in (24)-(26)

IF $\Delta\bar{x}_i > 0$: IF $\Delta\bar{\mu}_{U,i} > 0$: $\delta_{U,i} = 1$ and $\delta_{x,i} = \min[1, \Delta_{\rho U}]$

ELSEIF $\Delta\bar{\mu}_{U,i} < 0$: $\delta_{U,i} = \delta_{x,i} = \min[1, \Delta_{XU}]$

THEN $\delta_{L,i} = \min[1, \Delta_{\delta L}]$

ELSEIF $\Delta\bar{x}_i < 0$: IF $\Delta\bar{\mu}_{L,i} > 0$: $\delta_{L,i} = 1$ and $\delta_{x,i} = \min[1, \Delta_{\rho L}]$

ELSEIF $\Delta\bar{\mu}_{L,i} < 0$: $\delta_{L,i} = \delta_{x,i} = \min[1, \Delta_{XL}]$

THEN $\delta_{U,i} = \min[1, \Delta_{\delta U}]$

**Output:** $\delta_{x,i}$ and $\delta_{B\in\{L,U\},i}$

---

$$\Delta_{XB} = \frac{\bar{\Delta}_X - \bar{\Delta}_B - \sqrt{\bar{\Delta}_X^2 + (2\rho(\chi^k) - 1)2\bar{\Delta}_X\bar{\Delta}_B + \bar{\Delta}_B^2}}{2} \quad (24)$$

$$\Delta_{\rho B} = \frac{(\rho(\chi^k)\mu_B^k + \Delta\bar{\mu}_{B,i})}{(\mu_B^k + \Delta\bar{\mu}_{B,i})}\bar{\Delta}_X \quad (25)$$

$$\Delta_{\delta B} = \frac{(\rho(\chi^k)\bar{\Delta}_X - \delta_{x,i})}{(\bar{\Delta}_X - \delta_{x,i})}\bar{\Delta}_B \quad (26)$$

where $\bar{\Delta}_X$ and $\bar{\Delta}_B$ represents the recomputed feasible damping ratios from (19)-(20). Finally, the updated feasible NR states further ensure the hold of (21) obtained as:

$$x_i^{k+1} = x_i^k + \delta_{x,i}\tau_{x,i}\Delta x_i \quad (27)$$

$$\mu_{B,i}^{k+1} = \mu_{B,i}^k + \delta_{B,i}\tau_{B,i}\Delta\mu_{B_i} \quad (28)$$

### B. Optimal Residual Limiting: 3D-space search

From the perspective of representing a set of optimality conditions (10) as a network model, a vector of primal and dual variables $\mathbf{z}$ can be also seen based on its physical interpretability. Namely, $\mathbf{z} = [\mathbf{v}, \mathbf{g}, \mathbf{b}]^T$, where $\mathbf{v}$ corresponds to primal and dual variables representing network voltages, while $\mathbf{g}$ include variables related to impedance, as well as the real and reactive power variables respectively. This new perspective is further utilized to rethink a generic line search problem and transform it into a 3D space search problem,

$$\min_{\alpha,\beta,\gamma\in[0,1]}\|\theta(\mathbf{v}^k + \alpha\Delta\mathbf{v}, \mathbf{g}^k + \gamma\Delta\mathbf{g}, \mathbf{b}^k + \beta\Delta\mathbf{b})\|_2^2 \quad (29)$$

that utilizes the polynomial nature of network governing equations and constraints on its response to efficiently obtain a set of three damping factors ($\alpha, \beta, \gamma \in [0,1]$). Furthermore, it minimizes the problem residual at the current NR iteration.

The global solution to (29) that can be efficiently obtained using the methodology presented in [14], which minimizes the problem residual more or equal (same values of damping factors) as the solution of an exact line search problem. It also represents another example of how reaching beyond a generic set of algorithms while considering the nature of the problem can create a more efficient, problem-specific solution methodologies. Lastly, with consideration of the generality of other techniques, the 3D space search problem from (29) can be, without loss of generality, replaced with a form of traditional line search if another non-polynomial and or non-circuit based ACPF formulation is used.

### C. Variable Filtering technique

Thus far, this paper introduced a new set of NR step control techniques *independent of tuning parameters*. However, even with the vectorized NR step control, the nonlinear nature of ACPF model can, in general, cause small consecutive residual decrement, thereby requiring a step-size perturbation to prevent the efficiency issues and enable the future convergence. To that end, and in contrast to the generic, often used random step perturbation methods [3],[39], the Variable Filtering technique perturbs the NR step based on the physical characteristics of each variable obtained from the network representation in (10). Namely, as shown in (30), the variable perturbations are introduced by letting a NR step that remains within the "trusted" physical region $z_{th,i}$, "pass", and hence not get affected by damping factors that saturate the problem residual.

$$\mathbf{z}^k \to \begin{cases} z_i^{k+1} & \text{if } |\Delta z_i| < z_{th,i} \\ z_i^k & \text{if } |\Delta z_i| \geq z_{th,i} \end{cases} \quad (30)$$

With this Variable Filtering technique, the ECP algorithms can fully utilize the advantages of both circuit simulation and generalized optimization methods.

## V. Robust, Efficient and Scalable ACPF Analyses

In general, the components defining a robust, efficient, and scalable technology for analyzing and designing a power system steady-state response can be classified as:

1. Accurate and concurrent incorporation of all realistic, industry required control and operational steady-state constraints within an AC power flow model.
2. Scalability to realistic size, fully constrained models.
3. Robustness and efficiency of model simulation and building (optimization) methods.
4. Consideration of simplified, convex models *only* for initialization processes and cleaning the input data.

The lack of a generalized technology is projected to annually cost millions of dollars [40] and lead the operational and planning grid engineers towards basing their ACPF case "building" on simplified models. Moreover, these convexified power flow models are not AC feasible, and the transition to a fully feasible model can turn into days and even weeks and months long "guess and check" processes that require both experienced knowledge of grid physics, as well as the "art of tuning" of the traditional ACPF solvers [41]. Most importantly, the outcome potentially guarantees model solvability, but is often far away from an optimal point.

To address these real-life challenges, this section motivates a development and utilization of problem-specific physics-based numerical methods. Namely, it is demonstrated that the ECP framework allows for accurate and efficient AC power flow simulation and optimization processes while producing high quality solutions. Most importantly, even though the introduced set of physics-inspired algorithms cannot generally guarantee the global optimality of the processes, an efficiently obtained optimal solution that considers the realistic power grid constraints and operational limits is better than a non-optimal or even a global one that does not. Results for several examples are generated within a developed MATLAB prototype ECP-solver, on a MacBook Pro 2.9 GHz Intel i9.

### A. Realistic AC Power Flow simulations

To demonstrate generality as well as the benefit of exploiting the grid physics to obtain ACPF solutions efficiently and accurately, we first consider solving a set of ACPF problems defined in terms of equivalent circuit equations with implicitly constrained voltage regulation (VR), modeled with non-ideality factor $\varepsilon = 10^{-7}$, as given in (1)-(6). The set of cases solved within the prototype ECP-based solver, as well as generic MATLAB numerical solver FSolve, includes three synthetic ACPF models [29] corresponding to a 70k bus Eastern Interconnection (EI), a 25k bus Northeastern, and a 10k Western Interconnection (WECC) models. To provide a fair comparison between the physics-inspired and generic numerical methods, the *identical* initialization, together with the Jacobian matrix and function vector required to obtain NR-steps is passed to both solvers, thus making them *only* differ in the methods used for iterative NR-step control. The obtained convergence profiles are presented in Figure 10.

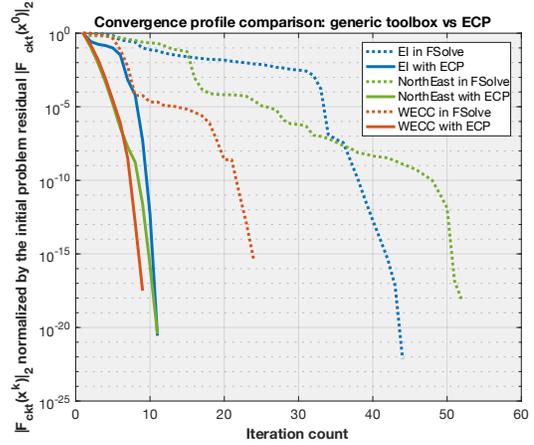

Figure 10. Comparing the convergence profiles of a generic numerical methodology vs. physics inspired ECP framework.

As expected, the utilization of the domain-specific knowledge within vectorized NR step control performed better as compared to a generic numerical toolbox. Notably, the difference in convergence profiles from Figure 10 may have meant a lot for the first computers that were used by those who first formulated power flow analyses on a computer, but today it "only" represents a nice result for publications, since both solvers converge within or less than a second. The obtained solution quality, however, represents a more valuable metric for the further comparisons. For example, consider the physical characteristic of a VR (voltage regulator) device from Figure 4 with the profile captured for all of the VRs in the Eastern Interconnection testcase, as shown in Figure 11.

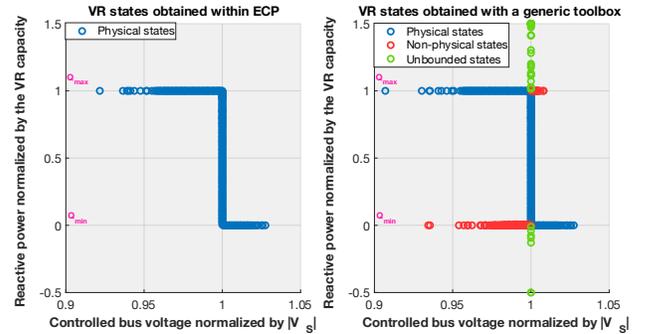

Figure 11. VR profiles for 5594 regulated system buses. Each VR state is obtained by normalizing the regulated bus voltage with respective setpoint $|V_s|$, and further plotting it as a function of reactive power supplied by the VR device. For plotting consistency, each reactive power value is shifted by its minimum operational limit and further normalized by its capacity.

In referring to both sets of 5594 VR states of 70k bus EI test case from Figure 11, it is important to note that all of them represent a solution to (5)-(6). However, *controlling each variable step based on the known physical characteristics* of VR devices (Figure 4) within ECP solver ensures the solution quality, i.e., the final solution corresponds to a physically realizable state. In contrast, not utilizing the knowledge of physical characteristics within a generic solver results in an ACPF solution with 404 VR states in the non-operational region, and an additional 52 states that do not respect the operational reactive power limits, which renders the ACPF solution obtained via a generic toolbox unusable. To that end, the lack of inclusion of the problem specific knowledge within the generic numerical methods may also serve as an explanation for the utilization of not always stable outer loop approaches [42] implemented to handle the disjunctive model behavior within the standard, industry accepted ACPF toolboxes.

Finally, with the efficient capability of implicitly enforcing a set of realistic operational and control characteristics on an ACPF model, the proposed framework naturally extends to include more accurate power flow optimizations.

### B. Realistic AC Power Flow optimizations

The AC-OPF problem that enforces the PQ capacity limits for generators as well as the bus voltage and thermal flow constraints on an ACPF model was first formulated over half of a century ago [43], but remains a challenging optimization problem [40]. We first analyze the performance of an AC-OPF problem defined and solved within ECP framework and compare it against the traditional formulations solved within generic optimization solvers in MATPOWER 7.0 [44]. The examined cases (Table 2), characterized in terms of Polar and Rectangular P/Q and I/V ACPF formulations, are solved using the effective known tools, KNITRO, FMINCON and MIPS [44], and compared with the solutions from our ECP solver. The MATPOWER best-case results based on the cost function values, followed by the iteration count, are presented in Table 2. To ensure that the compared approaches only differ in the formulation and numerical methods, the identical input file initialization is supplied to both solvers.

Table 2. AC-OPF iteration counts for the results obtained by KNITRO that performed the best among the three considered solvers, and the circuit formulation implemented within ECP framework. Note that both MATPOWER best-results and ECP solutions match in the optimal objective function values.

| case | # Bus | MATPOWER | | ECP | Cost [$/hr.] |
|------|-------|----------|-------------|--------|--------------|
| | | Iter. # | Formulation | Iter. # | |
| 1354pegase | 1,354 | 28 | Polar I/V | 15 | 74,064.2 |
| 2383wp | 2,383 | 32 | Polar P/Q | 22 | 1,863,597.5 |
| 2736sp | 2,736 | 23 | Polar P/Q | 19 | 1,307,998.3 |
| 2869pegase | 2,869 | 27 | Polar P/Q | 19 | 133,993.5 |
| 3012wp | 3,012 | 31 | Polar I/V | 24 | 2,584,033.9 |
| 6468rte | 6,468 | 35 | Polar I/V | 22 | 87,139.7 |
| 9241pegase | 9,241 | 56 | Polar I/V | 26 | 315,902.5 |
| ACTIVSg10k | 10,000 | 80 | Polar I/V | 25 | 2,488,650.0 |
| ACTIVSg25k | 25,000 | 48 | Polar P/Q | 24 | 6,019,821.2 |
| ACTIVSg70k | 70,000 | 138 | Polar P/Q | 34 | 16,538,287.9 |

As can be seen from Table 2, the polar formulations generally performed better within MATPOWER. However, the larger a problem gets, the more variables and inequality constraints, and therefore, the larger the impact of the conservatism that is introduced by the traditionally-used step limiting techniques. In contrast, the dependency between the problem size and the performance of the solution process is not as strongly indicated for the results obtained within the ECP solver, demonstrating the advantages of treating each of the NR step sizes more independently and based on their physical characteristics.

Next, consider a set of network models recently introduced for ARPA-E sponsored Grid Optimization (GO) Challenge 1 [45], given in Table 3, whose respective AC-OPF solutions are efficiently obtained within the ECP solver, as presented in

Figure 12. From the results shown in Table 3, it can be seen that none of the AC-OPF solutions are fully-contingency secured, which renders the respective optimal operating points practically unusable. The security of AC-OPF solutions can be ensured by representing each of the contingency events as an ACPF model given by (1)-(6) and appending its governing equations as an additional set of constraints to the originally defined AC-OPF problem. This is, however, easier said than done, since the problem size drastically increases with each additional contingency included. For instance, the network N13 with all 9519 contingencies results in a security constrained AC-OPF problem with ~1.6 billion variables, which represents an ideal example for testing the scalability of the introduced ECP concept.

Table 3. GO-Competition 1, Scenario 1 network models. The infeasible contingencies are identified by solving [38] an ACPF model given by (1)-(6) with obtained AC-OPF set points.

| Network | # Bus | # N-1 Cont. | # Infeasible Cont. |
|---------|-------|-------------|--------------------|
| N1 | 500 | 386 | 22 |
| N3 | 793 | 86 | 9 |
| N6 | 2000 | 2594 | 4 |
| N7 | 2312 | 953 | 34 |
| N8 | 3013 | 1959 | 3 |
| N81 | 3288 | 4650 | 73 |
| N84 | 4601 | 7075 | 48 |
| N9 | 4918 | 5065 | 17 |
| N12 | 9591 | 4377 | 62 |
| N13 | 10000 | 9519 | 70 |

To examine the scalability and impact the physics-inspired numerical methods have on the convergence of large-scale optimizations, we utilize the fully parallelizable Bordered-Block-Diagonal structure of the SC-AC-OPF problem and run it using a commercial version of SUGAR™ cloud-based platform (courtesy of Pearl Street Technologies) that is built on the introduced ECP methodology. The resulting iteration counts indicate a weak correlation between problem size and solution process efficiency are presented in Figure 12. Notably, the iteration counts are not significantly affected by the *number-of-contingency times* increase of problem sizes, thus demonstrating the scalability of the introduced ECP framework. Importantly, the results presented in Figure 12 indicate that the concept introduced in this paper can be applied to distributed optimization problems in general. In contrast to generic state-of-the-art approaches with a global NR-step damping methods, the localized nature of the NR-step limiting techniques can be applied per processor and does not require additional global communication.

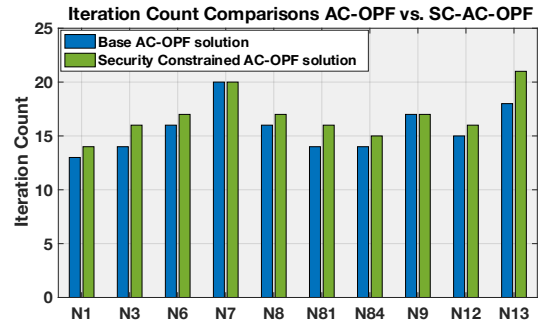

Figure 12. Physics-based vectorized NR step control ensures the weak correlation between the problem size and iteration count.

## VI. Conclusions

Due to climate change and the global warming crises, electrical power systems are evolving faster than ever before. To make an actual impact and solve the real-life challenges imposed on power system engineers it is necessary for the optimization formulations and methods to incorporate the physics from the grid. To that end, this paper postulates a novel concept for developing the problem specific numerical methods by infusing the knowledge of power system physics directly within the generalized optimization theory and algorithms. The introduced power grid-specific numerical methods were demonstrated to facilitate significant improvements in accuracy, robustness, and scalability of AC power flow analyses when compared against the existing generic "state-of-the-art" methodologies. Moreover, this approach is directly compatible with, and naturally extends to, distributed optimizations. The result is a foundation for analyzing and ensuring a reliable, sustainable, and resilient grid of the future.

## VII. References

[1] S. Boyd, L. Vandenberghe, *Convex Optimization*, Cambridge University Press, New York, NY, USA, 2004.

[2] E. Alpaydin, "Introduction to Machine Learning," 2nd Edition, MIT Press, 2010.

[3] Dimitri P. Bertsekas, "Nonlinear Programming," Athena Scientific, 2nd Edition, 1995.

[4] A. Das and B. K. Chakrabarti, "Quantum Annealing and Related Optimization Methods," Lecture Note in Physics, Vol. 679, Springer, Heidelberg (2005).

[5] W. Press, S. Teukolsky, W. Vetterling, B. Flannery, "Numerical Recipes in C – The Art of Scientific Computing," Second Edition, Cambridge University Press, 1992.

[6] S. Das and P. N. Suganthan, "Differential Evolution: A Survey of the State-of-the-art", IEEE Trans. on Evolutionary Computation, Vol.15, No.1, pp.4-31, Feb.2011.

[7] D. H. Wolpert and W. G. Macready, "No free lunch theorems for optimization," in IEEE Transactions on Evolutionary Computation, vol. 1, no. 1, pp. 67-82, April 1997. doi: 10.1109/4235.585893.

[8] B.D.H, Tellegen, "A general network theorem with applications," Philips Research Reports. 7: 259–269, 1952.

[9] Nagel, L. W, and Pederson, D. O., SPICE (Simulation Program with Integrated Circuit Emphasis), Memorandum No. ERL-M382, University of California, Berkeley, Apr. 1973.

[10] L. W. Nagel, "SPICE2: A Computer Program to Simulate Semiconductor Circuits," Memorandum No. ERL-M520, University of California, Berkeley, May 1975.

[11] L. Pileggi, R. Rohrer, C. Visweswariah, "Electronic Circuit & System Simulation Methods," McGraw-Hill, Inc, NY, USA, 1995.

[12] M. Celik, L. Pileggi. A. Odabasioglu, "IC Interconnect Analysis," Kluwer Academic Publishers, Springer, Boston, MA, 2002.

[13] Private communication with Hui Zheng from Pearl Street Technologies and Altan Odabasioglu from ANSYS.

[14] M. Jereminov, "Equivalent Circuit Programming," Doctoral thesis, Department of Electrical and. Computer Engineering, Carnegie Mellon University, Pittsburgh, PA, USA, August 2019.

[15] C. A. Desoer and E. S. Kuh, "Basic Circuit Theory," McGraw-Hill, New York, 1969.

[16] R. Byrd, J. Gilbert, and J. Nocedal, A trust region method based on interior point techniques for nonlinear programming, Math. Programming, 89, 149-185, 2000.

[17] R. Byrd, R.H., Mary E. Hribar, and Jorge Nocedal, "An Interior Point Algorithm for Large-Scale Nonlinear Programming," SIAM Journal on Optimization, Vol 9, No. 4, pp. 877–900, 1999.

[18] R. A. Waltz, J. L. Morales, J. Nocedal, and D. Orban, "An interior algorithm for nonlinear optimization that combines line search and trust region steps," Mathematical Programming, Vol 107-3, pp.391–408, 2006.

[19] "The disaster that could follow from a flash in the sky," Economist, July 13, 2017.

[20] D. Bromberg, M. Jereminov, L. Xin, G. Hug, L. Pileggi, "An Equivalent Circuit Formulation of the Power Flow Problem with Current and Voltage State Variables," PowerTech Eindhoven, June 2015.

[21] M. Jereminov, D. M. Bromberg, L. Xin, G. Hug, L. Pileggi, "Improving Robustness and Modeling Generality for Power Flow Analysis," T&D Conference and Exposition, 2016 IEEE PES.

[22] A. Pandey, M. Jereminov, M. Wagner, D. M. Bromberg, G. Hug, L. Pileggi, "Robust Power Flow and Three Phase Power Flow Analyses", IEEE Trans. on Power Systems. DOI: 10.1109/TPWRS.2018.2863042.

[23] A. Pandey, M. Jereminov, G. Hug, L. Pileggi, "Improving power flow robustness via circuit simulation methods," IEEE PES GM, 2017.

[24] M. Jereminov, A. Pandey and L. Pileggi, "Equivalent circuit formulation for solving AC optimal power flow," IEEE Transaction on Power Systems, DOI:10.1109/TPWRS.2018.2888907.

[25] A. Pandey, M. Jereminov, M. R. Wagner, G. Hug and L. Pileggi, "Robust Convergence of Power Flow using Tx Stepping Method with Equivalent Circuit Formulation" in 2018 (PSCC), Dublin 2018.

[26] M. Jereminov, A. Terzakis, M. Wagner, A. Pandey, L. Pileggi, "Robust and Efficient Power Flow Convergence with G-min Stepping Homotopy Method," in Proc. IEEE Conference on Environment, Electrical Engineering and I&CPS Europe, Genoa, Italy, June 2019.

[27] W. F. Tinney and C. E. Hart, "Power flow solutions by Newton's method," IEEE Trans. on PAS, Vol. 86, No. 11, pp. 1449-1460, 1967.

[28] H. W. Dommel, W. F. Tinney, and W. L. Powell, "Further developments in Newton's method for power system applications," in Proc. IEEE Winter Power Meeting Conf., New York, NY, USA, Jan. 1970.

[29] A. B. Birchfield, et.al., "Power Flow Convergence and Reactive Power Planning in Creation of Large Synthetic Grids", IEEE Trans. on Power Systems, 2018.

[30] A. F. Glimn, G.W. Stagg, "Automatic Calculation of Load Flows," AIEE Summer General Meeting, Montreal, Canada, June 1957.

[31] J. B. Ward, H. W. Hale, "Digital Computer Solution of Power Flow Problem," AIEE Winter General Meeting, NY, USA, June 1956.

[32] H. H. Happ, "Piecewise Methods and Applications to Power Systems," John Wiley and Sons, New York. 1980.

[33] W. Murray, T. T. De Rubira, and A. Wigington, "Improving the robustness of Newton-based power flow methods to cope with poor initial conditions," North American Power Symposium (NAPS), 2013.

[34] A. G. Exposito, E. R. Ramos, "Reliable Load Flow Technique for Radial Distribution Networks," IEEE Transactions on Power Systems, Vol. 14, No. 3, August 1999.

[35] P. A. N. Garcia, J. L. R. Pereria, S. Carneiro Jr., M. P. Vinagre, F. V. Gomes, "Improvements in the Representation of PV Buses on Three-Phase Distribution Power Flow," IEEE Transactions on Power Systems, Vol. 19, No. 2, April 2004.

[36] L. Chua, C. Desoer, E. Kuh, "Linear and Nonlinear Circuits," McGraw–Hill Companies, 1987.

[37] S.W. Director, R. Rohrer, "Automated Network Design-The Frequency Domain Case", IEEE Trans. on Circuit Theory, vol. 16, no3, August 1969.

[38] M. Jereminov, D. M. Bromberg, A. Pandey, M. R. Wagner, and L. Pileggi, "Evaluating Feasibility Within Power Flow," in IEEE Transactions on Smart Grid, vol. 11, no. 4, pp. 3522-3534, July 2020, DOI: 10.1109/TSG.2020.2966930.

[39] N. Tripuraneni, M. Stern, C. Jin, J. Regier, and M. I. Jordan, "Stochastic cubic regularization for fast nonconvex optimization," In Proceedings of the 32nd International Conference on Neural Information Processing Systems (NIPS'18), NY, USA, 2904–2913. 2018.

[40] M.B. Cain, R. P. O'Neill, and A. Castillo, "History of optimal power flow and formulations, OPF Paper 1" Tech. Rep., US FERC, Dec. 2012.

[41] D. Bromberg, A. Pandey, H. Zheng, Y. Li, "Robust Solution of High Renewable Penetration Planning Cases in SUGAR," Technical Conference on Increasing Real-Time and Day-Ahead Market Efficiency and Enhancing Resilience through Improved Software, FERC, 2019.

[42] J. Katzenelson, "An algorithm for solving nonlinear resistor networks," in The Bell System Tech. Journal, vol. 44, no. 8, pp. 1605-1620, Oct. 1965.

[43] J. Carpentier, "Contribution to the economic dispatch problem," Bulletin de la Societe Francaise des Electriciens, 8 (1962), pp. 431–447.

[44] R. Zimmerman, C. Murillo-Sanchez, R. Thomas, "MATPOWER: Steady-state operations, planning and analysis tools for power systems research and education," IEEE Trans. on Power Systems, vol. 26, no.1, Feb 2011.

[45] https://gocompetition.energy.gov/challenges/22/datasets.